# Depth from Defocus Technique Applied to Unsteady Shock-Drop Secondary atomization

Shubham Sharma[1*], Saini Jatin Rao[1], Navin Kumar Chandra[1], Aloke Kumar[1], Saptarshi Basu[1,3] and Cameron Tropea[2*]

[1]Department of Mechanical Engineering, Indian Institute of Science - Bangalore, Bengaluru, 560012, India.
[2]Institute of Fluid Mechanics and Aerodynamics, Technical University of Darmstadt, Darmstadt, 64287, Germany.
[3]Interdisciplinary Centre for Energy Research, Indian Institute of Science - Bangalore, Bengaluru, 560012, India.

*Corresponding author(s). E-mail(s): shubhams12@iisc.ac.in; ctropea@sla.tu-darmstadt.de; Contributing authors: jatinrao@iisc.ac.in; navinchandra@iisc.ac.in; alokekumar@iisc.ac.in; sbasu@iisc.ac.in;

**Abstract**

The two-sensor depth from defocus (DFD) technique for the measurement of drop sizes in a spray is further developed to achieve higher spatial and temporal resolution, to improve estimates of size and number concentration, and to provide additional guidelines for the calibration and design of the optical system for a specific application. The technique and these improvements are demonstrated using the case of secondary atomization when a shock wave interacts with a single drop. This is an application in which both high spatially and temporally resolved number density and size distributions of secondary droplets generated in the wake of the original drop are necessary.

# 1 Introduction

This study introduces further developments of the depth from defocus (DFD) imaging technique, used to measure the size and position of liquid drops in a spray. The technique is demonstrated using the case of secondary atomization, which occurs in a large number of atomization processes[1, 2]. Whereas primary atomization involves the first division of fragments from a bulk liquid [3], secondary atomization occurs where and when the resulting liquid fragments and drops are again broken into even smaller droplets[4]. It is therefore the secondary atomization which determines the final drop size distribution issuing from an atomiser and it is this distribution which often defines the performance and/or efficiency of the atomising process. Numerous examples of the importance of the secondary atomization can be given; however, the present study focuses on one particular example, namely the aerobreakup of drops and its measurement, in particular in high-speed flows. Specifically, when producing metal powders for laser bed fusion additive manufacturing, molten metal is atomised using a close-couple atomiser[5]. In this atomiser, a gravity fed molten metal stream is atomised using a concentric supersonic gas jet, typically of inert gas. Flow Mach numbers exceeding 5-6 are easily achieved.

In predicting the drop size distribution of such an atomiser, the aerobreakup of liquid fragments and drops is therefore an important process to model, since this is an essential element in determining the resulting metal powder size. Numerous modelling schemes have been proposed for this purpose, generally subdivided according to the local Weber number acting on the drop, defined using the relative velocity of the drop compared to the flow (slip velocity), $U$, the gas density, $\rho_g$, the initial drop diameter, $D_0$, and the liquid to gas surface tension, $\sigma$: We= $\rho_g U^2 D_0 / \sigma$. Typical breakup modes include vibrational (0 < We < 11), bag breakup (11 < We < 35), multimode breakup (35 < We < 80), sheet-thinning or shear stripping (80 < We < 350), and catastrophic breakup (We > 350) [4, 6].

Preliminary visualisation studies of shock-drop interaction have indicated that the breakup of single drops following the interaction with a shock wave can involve several different breakup modes, due in part to the varying relative velocity in time, but also due to the sharply changing morphology of the drop during breakup[7]. This observation underlines the necessity to measure not only the overall size distribution resulting from the shock-drop interaction, but to monitor the secondary droplet size in time and space. Only then can the measured size be representative of the instantaneous breakup mode and thus, be of use for predictive model formulation. This requires therefore a transient measure of droplet size distribution over a volume, and the purpose of this article is to describe a suitable technique for performing such a measurement. The technique used is based on a depth from defocus (DFD) imaging, with which a transient measurement of size and number of drops within a welldefined control volume can be performed[8]. Specifically, further developments have been introduced to this technique to increase the overall accuracy and to better define the measurement volume, resulting in improved estimates of size distributions and number density.

This article is structured in the following manner. Experimental details of the shock-drop facility are presented in section 2. The details of the DFD technique and its implementation for measuring secondary drop sizes are explained in section 3. The results are provided in section 4 and the conclusions in section 5.

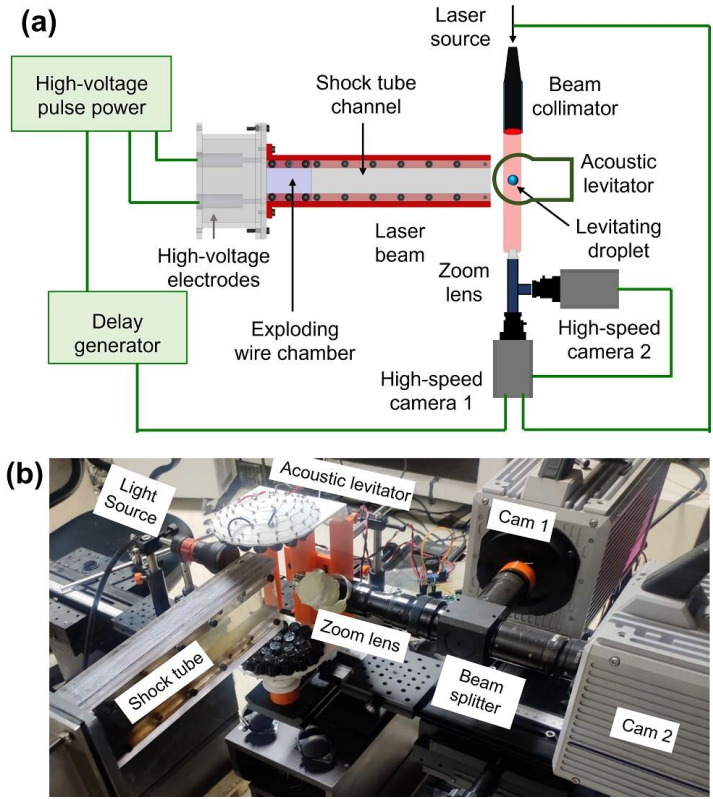

**Fig. 1** Wire-blast based shock tube setup. (a) Schematic and (b) photograph of the depth from defocus optical arrangement, including the shock tube and acoustic levitator.

## 2 Experimental Set-up

A model problem of shock-drop interaction is chosen in the present work to demonstrate the ability of the DFD technique to provide sizing information of a complex phenomenon requiring high spatio-temporal resolution. In this process, a primary drop is held in the path of an oncoming normal shock wave, resulting in its fragmentation into numerous daughter droplets. This interaction, specifically at high Weber numbers ($\sim O(10^3)$), results in the formation of a dense mist of micron-sized daughter droplets (requiring high spatial resolution) which evolves over time. Furthermore, the interaction dynamics are complete in an extremely short time ($\sim O(100)\mu s$), thus necessitating high temporal resolution. A detailed description of the experimental setup used for achieving shock-drop interaction can be found in previous work [6, 7, 9]; however, the essential details of the test facility are also described here. The arrangement of equipment in a shock-droplet interaction setup is shown in Fig. 1. Figure 1a shows the schematic representation of the electrical wiring diagram for different components, while Fig. 1b shows the photograph of DFD optical arrangement. The generation of a normal shock wave is achieved by using an exploding wire based shock tube setup. This involves an explosion of a copper wire (35 SWG, bare copper wire) by passing a high current and high voltage electrical energy pulse through it for a very short time duration ($\sim O(1)\mu s$). It results in the formation of a blast wave, which is transformed into a planar shock wave due to the constricting rectangular geometry (20 mm x 50 mm x 320 mm) of the shock tube channel (see Fig. 1). A 2kJ pulse power system (Zeonics Systech India, Z/46/12) provides electrical energy for wire explosion. An acoustically levitated de-ionized (DI) water droplet of size $D_o \sim$ 2.5 mm is held axially at a distance of 15 mm from the exit of the shock tube channel using a multi-transducer, single-axis acoustic levitator (Big-LEV, [10]). A simultaneous trigger signal is provided to the imaging system and high voltage pulse power system to initiate the experimental trial using a digital delay generator (BNC 575).

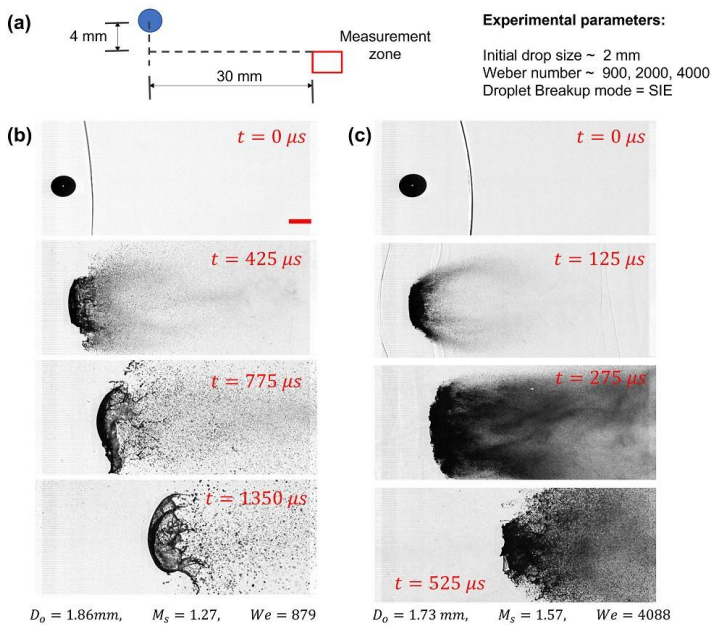

**Fig. 2** : Global observation of drop aerobreakup and resulting daughter droplets produced. (a) $D_o$ = 1.86 mm, $M_s$ = 1.27, $We$ = 879, (b) $D_o$ = 1.73 mm, $M_s$ = 1.57, $We$ = 4088. Scale bar is equal to 2 mm.

Details of the imaging system employed for DFD measurements are provided in section 3. The global observation of the interaction phenomenon is obtained by using the experimental arrangement mentioned in previous work [9]. A sample observation of the global view of interaction phenomenon for $We \sim$ 900 and 4000 is shown below in figure 2. The global observations are also utilised for measuring shock Mach number ($M_s$) and Weber numbers (*We*) [7, 9]. Different values of the shock Mach number ($M_s$ = 1.27,1.39 and 1.56) and Weber number (We ∼ 900,2000 and 4000) are achieved by changing the electrical energy input to the exploding wire, while the diameter of the levitated drop was left unchanged.

# 3 Depth from defocus technique for drop size measurement

## 3.1 Measurement principle

In choosing a measurement technique for sizing the secondary droplets created from the shock-drop interaction, an imaging approach was highly desirable, since techniques only detecting droplets at a single point, like for instance the phase Doppler or time-shift technique[11], have the disadvantage that spatial variations of drop sizes can only be registered by traversing the measurement point and measuring over a very large number of single experiments. This is far too tedious, considering the preparation and run time of each experiment, as described above in section 2. The ILIDS/IPI technique[12, 13] allows droplets within an entire illuminated plane to be sized simultaneously, but then the illumination plane must be traversed throughout the cloud of droplets, again leading to a large number of experimental repetitions. Of all the techniques remaining, only holography and direct imaging are therefore viable options for this measurement task[14–17]. In either case a large depth of field is desirable; however, holography requires a coherent, polarised light source and in general is more complicated in setup, adjustment and operation. Also, holography requires high computational power for numerical reconstruction of recorded holograms and has uncertainty in the particle depth location [17]. Furthermore, to achieve the required time resolution high-speed cameras must be used, which then sacrifice resolution for speed - not desirable for digital holography. For these reasons, direct imaging was chosen for this study, whereby the large measurement volume in depth was achieved using the depth from defocus technique.

Following the first introduction of the depth from defocus technique[18, 19], several implementations were demonstrated, based either on single sensor[20, 21] or two sensor systems[22, 23]. In the present study a two-camera configuration is used with some modifications from previously employed systems[8, 24].

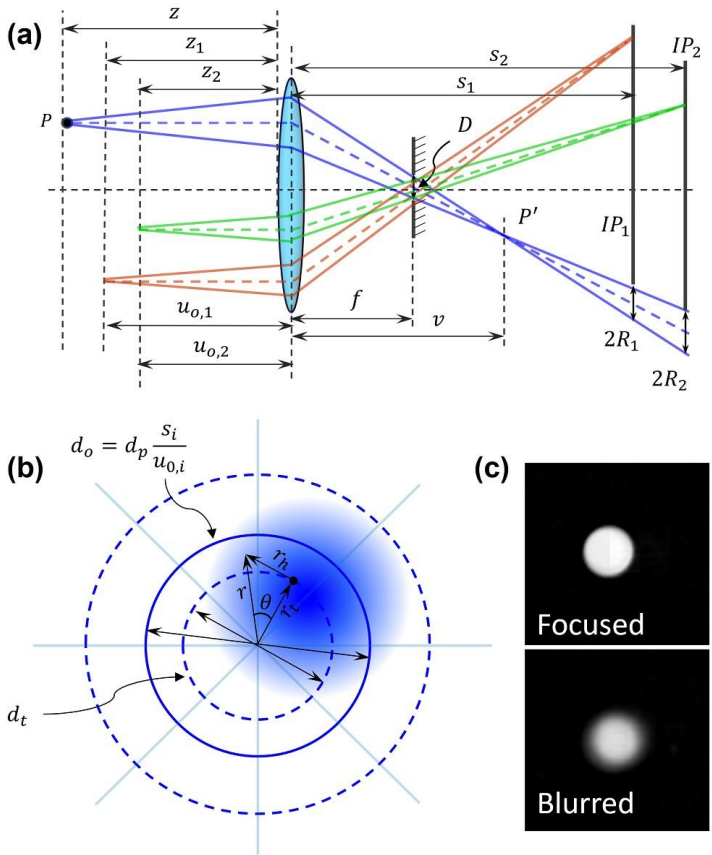

**Fig. 3** Depth from defocus technique (a) An object (e.g. drop) in the $z_i$ plane is in focus in the image plane ($IP_i$) at the distance $s_i$ from the lens. Here, $i$ = 1,2 corresponds to camera 1 and camera 2. Objects in front or behind the object plane ($z_i$) appear blurred on the image plane. (b) The amount of blur is computed by convolving the focused image with a blur kernel, shown as a shaded circle. Here, $u_{o,i}$ is the object plane distance from the lens optical axis. (c) The resulting images on the camera exhibit different degrees of blur, depending on the $z$ position of the drop.

The working principle is now briefly outlined, whereby further details can be found in the above cited articles.

With reference to Fig. 3, an object at some distance $z$ from a lens ($u_{o,i}$ from the optical axis of lens) on the object plane $P$ is focused on the image plane $P'$ according to the Gaussian thin lens equation

$$\frac{1}{f} = \frac{1}{u_{o,i}} + \frac{1}{v} \qquad M = \frac{v}{u_{o,i}} \qquad (1)$$

where $f$ is the focal length of the lens and $M$ is the transverse magnification.

Objects in front of or behind the object plane appear then defocused on the image plane with a blur degree dependent on the $z$ position. With no blur or diffraction a particle/drop of diameter $d_p$ will appear on the image plane with diameter $d_0 = M d_p$. The image intensity ($g_t$) at any location $r_t$ from the particle centre can be obtained by taking the convolution of focused image, $f(r)$, with a blur kernel, $h(r)$:

$$g_t(r_t) = f(r) * h(r_t - r) \qquad (2)$$

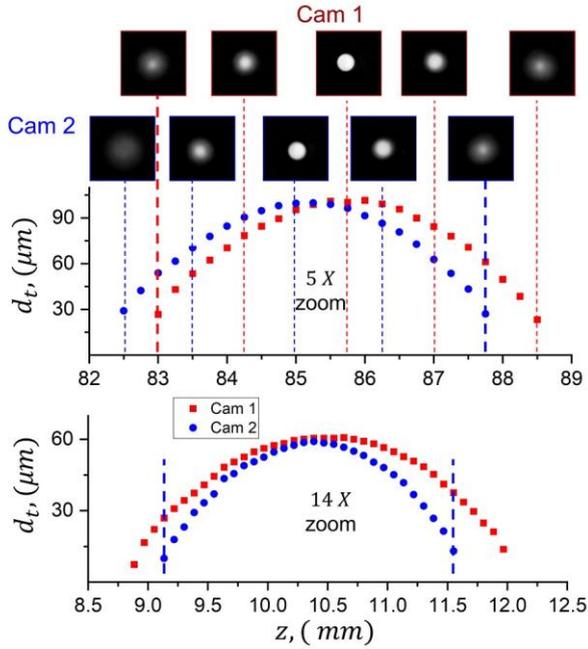

**Fig. 4** Calibration of the blurred images showing the resulting functions $k_i$ in red and blue for the two cameras and two different magnifications. $d_t$ is the image diameter at a grey level of $g_t$=0.6.

The normalised intensity of a focused image of a particle with image radius $r_o$ can be represented as:

$$f(r) = 0, \quad \text{if} \quad r > r_o$$
$$\phantom{f(r)} = 1, \quad \text{if} \quad 0 < r < r_o \qquad (3)$$

The blur kernel $h(r)$ can be represented using a Gaussian profile with $\sigma$ as the standard deviation [8, 25]:

$$h(r) = \frac{1}{2\pi\sigma^2} e^{-\frac{(\vec{r}-\vec{r_t})^2}{2\sigma^2}} = \frac{1}{2\pi\sigma^2} e^{-\frac{(r^2 + r_t^2 - 2rr_t \cos\theta)}{2\sigma^2}} \qquad (4)$$

Therefore, the two-dimensional convolution Eq. (2) can be written as:

$$g_t(r_t) = \int_0^{2\pi} \int_0^{d_o/2} \frac{1}{2\pi\sigma^2} e^{-\frac{(r^2+r_t^2-2rr_t\cos\theta)}{2\sigma^2}} \, rdr \, d\theta \tag{5}$$

The standard deviation of the blur kernal is given as

$$\sigma = Ad_p \frac{|\Delta z|}{z-f} \tag{6}$$

or

$$\frac{\sigma}{d_o} = \frac{ADM}{2f} \frac{|\Delta z|}{d_o} = \beta \frac{|\Delta z|}{d_o} \tag{7}$$

Here *A* is an experimental constant dependent on the imaging system, *D* is the aperture diameter (see Fig. 3 (a)), and Δ*z* is the distance of the drop from the object plane. As the values of *A, D, M* and *f* are fixed during DFD measurements; these terms are replaced with a single constant *β*. Thus, the size of the drop image ($d_t$), given a prescribed grey scale value ($g_t$ - blur degree), is a function *k* of the actual size of the drop ($d_p$) and its position along the optical axis (Δ*z*). This provides one equation for two unknowns; however, if two cameras with two different degrees of out-of-focus are used, i.e. with two different *k*-functions, then two equations for the two unknowns, $d_p$ and Δ*z*, are available:

$$\frac{d_{t,i}}{d_p} = k_i \left( \frac{\Delta z_i}{d_p} \right) \tag{8}$$

The calibration functions $k_i$ can be established by moving a target plate with different sized dots along the *z* axis, resulting in images and functions as pictured in Fig. 4. In this figure the calibration function *k* is shown for two different magnifications and the image size $d_t$ is taken at the grey value of $g_t$ = 0.6, whereby the image grey levels have been normalised between 0 (black) and 1 (white), as expressed in Eq. (3). Note that a telecentric lens has been used to ensure that the magnification factor remains insensitive to the exact position of the drop along the *z* axis. Further details regarding the choice of optical parameters can be found in Ref. [24]. A photograph of the optical arrangement, including the shock tube and acoustic levitator is shown in Fig. 1.

## 3.2 System realisation

In the present system the light source was a Cavitar cavilux smart UHS pulsed laser connected to a beam expander (Thorlabs GBE05-A) via fiber optics. The expanded beam was synchronised with the two Photron SA5 cameras. The cameras were operated at a frame rate of 20,000Hz with 576 x 624 pixels. The effective pixel resolution in the object plane depended on the magnification used in the 6.5x Navitar zoom lens coupled with 2x objective or Navitar 12X Ultra zoom lens coupled with 4x microscopic objective (Olympus Plan 4X), but was either 3.90*μ*m/pixel or 1.41*μ*m/pixel. A beam splitter was used to image the same field of view onto the two cameras. The difference in out-of focus degree was achieved by adding a spacer in front of one of the cameras. The alignment of the two fields of view was performed using the MATLAB® Registration Estimator App function, an approach which differs from previous procedures in that it is not based on aligning only circular shapes, but can work with arbitrary shapes by identifying and mapping distinguishable features using a feature-based alignment technique, namely a Speeded-Up Robust Features (SURF) algorithm. Although the present work assumed that all drops were spherical (circular in the image plane), in principle the DFD arrangement can be used for sizing also non-spherical liquid fragments. In that case however, the convolution integral yielding the analytic expression for the blurred image is no longer available in closed form, but must be computed numerically. Furthermore, the necessity would arise to make estimates of the size of the three-dimensional fragment from two-dimensional descriptors thereof, obtained from the projection of the object onto the imaging plane.

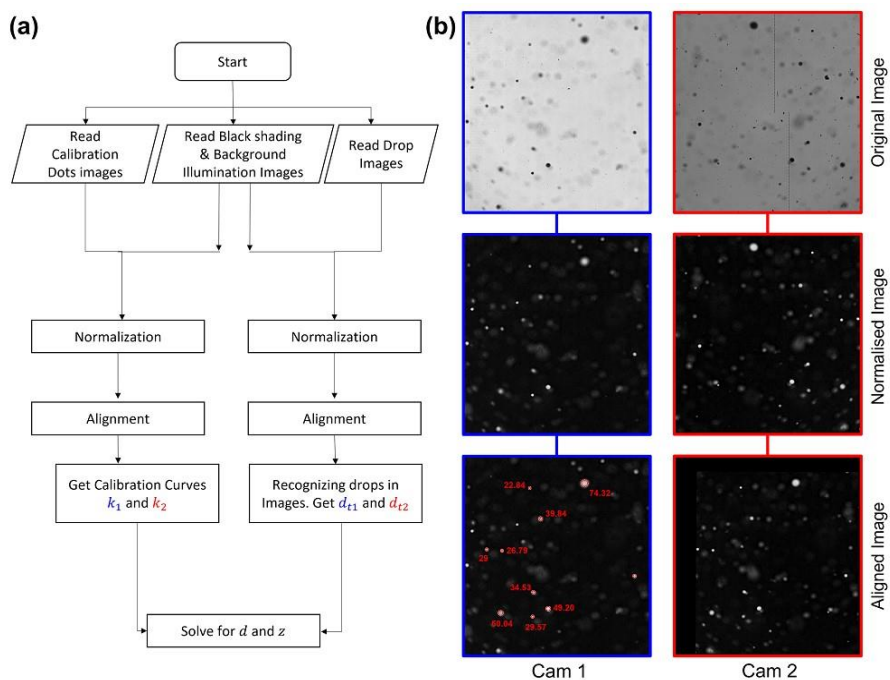

**Fig. 5** (a) Flowchart of image processing steps for the DFD. (b) Corresponding images: original images, normalised images and aligned images. The detected and sized drops from these images are marked with red circles in the final image.

The processing of the images follows the flowchart pictured in Fig. 5a, in which first the calibration curves $k_1$ and $k_2$ are obtained using a dot target (left side) and when measuring, the right side of the flow chart is followed. The results of the normalisation and alignment steps are pictured in Fig. 5b for two typical and simultaneous frames from the two cameras. The final detected and sized drops are shown as red circles in the final photograph.

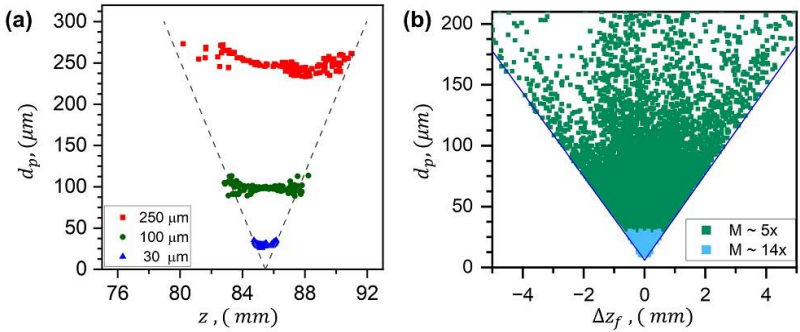

**Fig. 6** (a) Measured detection depth $z$ for three different size particles. This data was collected using dot target plates and the object plane was located at $z = 85.5$mm. (b) Sampled atomised secondary droplet diameters $d_p$ and corresponding depth from object plane $\Delta z_f$, here particles with $d_p > 30 \mu m$ considered from ~5x magnification configuration and particles with $d_p < 30 \mu m$ from ~14x configuration matched with suitable conversion factor.

### 3.3 Detection volume

One of the distinguishing features of the two-camera DFD arrangement is the well-defined volume over which drops are detected and measured. The detection depth $\delta$ of blurred drop images, given a constant grey scale detection level $g_t$, is linearly dependent on the size of the drop. Thus, larger drops are detected and sized over a larger range of $z$ than smaller drops. This is illustrated in Fig. 6a, in which the detection depth is shown for three different sized dots on the calibration target plate, which was traversed along the optical axis in front of and behind the object plane. The linear relation between particle size and detection depth of field is evident in Fig. 6a and b and can be quantitatively expressed as $\delta = 2\alpha(d_p - d_{p,0})$, where $\alpha$ is the inverse slope of the dotted lines in Fig. 6 and $d_{p,0}$ is a parameter to adjust the linear fit of these lines.

The detection depth $\delta$ for each drop size is estimated theoretically using Eq. (5). For the limiting case, when a droplet lies at the farthest $\Delta z$ location (i.e. at $\delta/2$), where it can just be detected on the camera sensor for the chosen threshold intensity value $g_{tc}$ (critical grey scale level), the detected image radius $r_t \to 0$. Eq. (5) for this limiting case yields:

$$g_{tc} = \int_0^{2\pi} \int_0^{d_o/2} \frac{1}{2\pi\sigma^2} e^{-\frac{r^2}{2\sigma^2}} r \, dr \, d\theta \tag{9}$$

or

$$g_{tc} = 1 - e^{-\frac{1}{8}\left(\frac{d_o}{\sigma}\right)^2}$$

(10)

and using Eq. (7) the detection depth is estimated as:

$$\delta = \frac{2f}{AD\sqrt{\ln\frac{1}{(1-g_{tc})^2}}} d_p \approx 2\alpha d_p$$

(11)

This suggests a linear relationship between detection depth $\delta$ and the drop size $d_p$, as observed in Fig. 6a. The linear variation is also validated in Fig. 6b, which shows detected drop sizes during all measurements taken in the present work. The above theoretical estimation of $\delta$ is obtained by considering an idealised case where $r_t \to 0$ when $g_t = g_{tc}$, as shown in Fig. 7a and c. However, in practical situations, the minimum detection of image size equals the pixel size ($s_p$) of the camera sensor (see Fig. 7b). Therefore, $r_{tmin} = s_p/2$ and

$$\tilde{\rho}_{t,m} = \frac{r_{t,min}}{d_o} = \frac{s_p}{2Md_p}$$

(12)

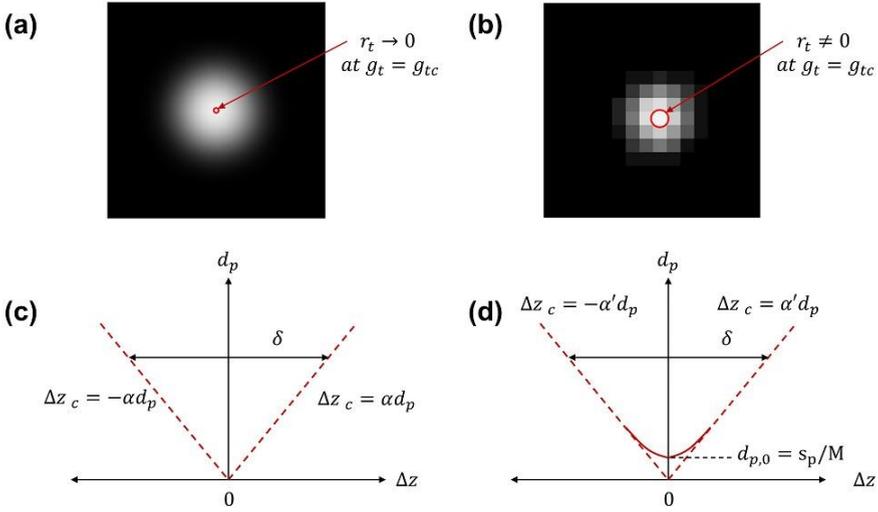

**Fig. 7** Estimation of the detection depth. (a,c) Ideal theoretical estimate of detection depth, where $r_t \to 0$ at $g_t = g_{tc}$, (b,d) theoretical estimate of detection depth by incorporating the camera sensor pixel size ($s_p$) as minimum detection length.

If the drop size is already known, $\tilde{\sigma}$ can be evaluated using iso-contour lines of $g_t$ (see Fig. 10a) and can be accurately fitted using the following function form for $g_t > 0.5$

$$|\tilde{\sigma}| = a\left(1 - e^{p(\tilde{\rho}_t - 0.5)}\right)$$

(13)

where $a$ and $p$ are the fitting parameters proportional to the image intensity value ($g_t$). Using Eqs. (7) and (12), the detection depth can be obtained as:

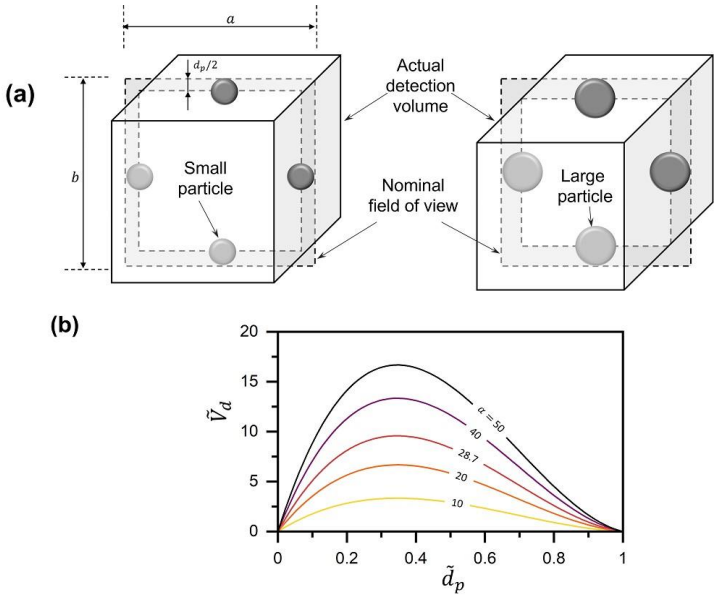

**Fig. 8** (a) Graphical illustration of detection volume size, showing size dependence of detection depth and field of view; (b) Dimensionless detection volume $\tilde{V}_d$ as a function of particle diameter for the conditions $\alpha$ = 28.72m/m, $a$ = 2246μm, $b$ = 2433μm, $\gamma$ = 1.083. and $d_{p,0} \approx$ 5μm.

$$\delta = \frac{2fa}{AD}\left(1 - e^{p\left(\frac{s_p}{2Md_p} - 0.5\right)}\right) d_p \tag{14}$$

This variation of detection depth with drop size is shown in Fig. 7d. This suggest that the $d_{p,o}$ is the diameter of the smallest detectable particle, i.e. the particle which first exceeds the $g_t$ threshold in the image. For a limiting case when $r_{t,min} \ll d_0$ i.e. for larger particles in comparison to $s_p/2M$, $\tilde{\rho}_{t,min} \ll 0.5$, in which case Eq. (14) can be reduced to

$$\delta = \frac{2fa}{AD}\left(1 - e^{-p/2}\right) d_p = \alpha' d_p \tag{15}$$

Thus, the linear relationship between $\delta$ and $d_p$ (as shown in Fig. 6 and Eq. (15)) is reproduced for larger drop sizes. If a certain size particle exists outside the $\delta$ limits shown in this figure, then its grey scale $g_t$ no longer exceeds the prescribed threshold value (in this work $g_t$ = 0.6); thus, the image size $d_{t,i}$ cannot be determined. As evident from Eq. (15) the inverse slope $\alpha'$ is determined by a number of factors, including the illumination, the optical configuration and the detector sensitivity, but also by the choice of grey level threshold $g_t$. In the present system $\alpha \approx$ 28.72m/m, i.e. the detection depth increases by 2×28.72 = 57.44μm for a 1μm increase in drop size. In the system presented in [8] a value of $\alpha \approx$ 36.59m/m was found.

However, the detection volume of the system is not only determined by the $\delta$ depth over which drop images can be detected and evaluated, but also by the field of view (FOV). The field of view is influenced by the particle size, since particles, whose images are truncated on the sides of the field of view, will not be processed. Hence, the field of view is reduced on all sides by the corresponding particle size. If the field of view is considered to be a rectangle of side lengths $a$ and $b$ in the object plane, then the effective field of view will have side lengths $a-d_p$ and $b-d_p$, as shown pictorially in Fig. 8a for a smaller, compared to a larger particle.

The detection volume can then be expressed as

$$V_d = 2\alpha(d_p - d_{p,0})(a - d_p)(b - d_p) \tag{16}$$

This detection volume is also shown in in Fig. 8a. The inverse of this volume

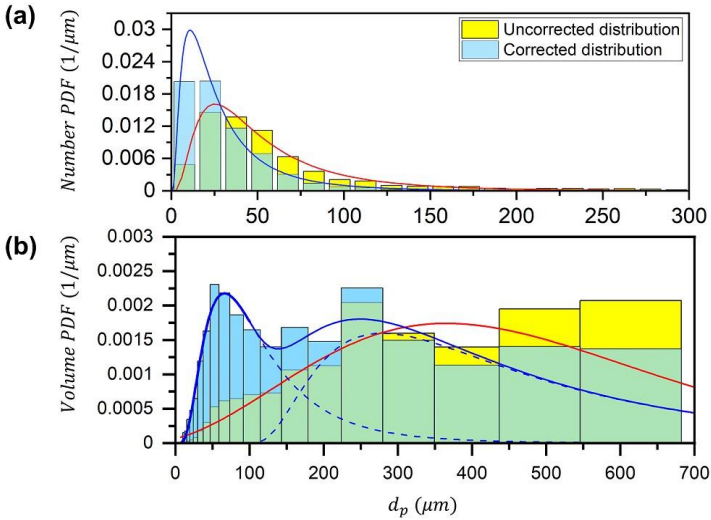

**Fig. 9** (a) Probability density function (PDF) of drop size and (c) volume weighted PDF, in which uncorrected and corrected estimates are shown. Data collected for We=900. The solid lines indicate the probability distribution function as per fitted log-normal function. For the volume distribution a multi-modal distribution was fitted.

must be used to weight the number of detected drops for each size to ensure an unbiased estimate of number density (drops per unit volume).

The detection volume $V_d$ is rendered dimensionless by first expressing the field of view in terms of the aspect ratio, i.e. $b = \gamma a$, then scaling lengths with $a$, and finally assuming $d_{p,0}/a \ll 1$, yielding:

$$\tilde{V}_d = \frac{V_d}{a^3} \approx \alpha\left[\gamma\tilde{d}_p - (1+\gamma)\tilde{d}_p^2 + \tilde{d}_p^3\right] \tag{17}$$

where the tilde represents dimensionless quantities. This dimensionless detection volume magnitude is pictured in Fig. 8b as a function of $\tilde{d}_p$ using a value of $\gamma$ typical for the present optical configuration ($\gamma$ = 1.083). Several values of $\alpha$ have been shown on this diagram. The figure indicates that the detection volume increases approximately linearly over a large range of drop/particle sizes, but levels off and reaches a maximum when the drop size approaches about 1/3 the field of view side length $a$. The maximum detection volume can be analytically computed by setting $d\tilde{V}_d/d\tilde{d}_p$ = 0, resulting in $\tilde{V}_{d,max}$ at $\tilde{d}_{p,max} \approx$ 1/3.

The effect of this weighting is exemplary illustrated in Figs. 9a and 9b in which the uncorrected and corrected probability density functions of the drop size and drop volume are shown respectively. As can be seen from the diagrams, the number of small drops (and their volume contribution) increases significantly after applying the detection volume correction. This is one of the key improvements implemented in the present work and presently, there is no alternative measurement technique for drops in a spray which offers comparable accuracy in estimating the detection volume size.

## 3.4 Improved calibration procedures

This section will introduce a simplified and improved approach for obtaining the calibration curve plotted in Fig. 4, which is obtained after imaging a target dot of known size at incremental locations along the optical axis. This requires a precise and gradual movement of the target plate at numerous locations, which is cumbersome and could also lead to measurement errors if a precise mechanism for target plate movement is not employed. Here a novel technique for theoretically obtaining the calibration curve is proposed, using a blurred image of a known size target dot kept at a single known distance from the object plane.

Using appropriate substitutions and integrating over $\theta$, Eq. (5) can be reduced to:

$$g_t(\tilde{\rho}_t) = \frac{1}{\tilde{\sigma}^2} \int_0^{1/2} e^{-\frac{\left[\left(\frac{\tilde{\rho}}{\tilde{\sigma}}\right)^2 + \left(\frac{\tilde{\rho}_t}{\tilde{\sigma}}\right)^2\right]}{2}} I_o\left(\frac{\tilde{\rho}\tilde{\rho}_t}{\tilde{\sigma}^2}\right) \rho d\rho \qquad (18)$$

where $I_o$ is a zeroth order modified Bessel function of the first kind and with the dimensionless quantities $\tilde{\rho} = \frac{\rho}{d_o}$; $\tilde{\rho}_t = \frac{\rho_t}{d_o}$; $\tilde{\sigma} = \frac{\sigma}{d_o}$. For different values of $\tilde{\rho}_t$ and $\tilde{\sigma}$ a contour plot for $g_t(\tilde{\rho}_t)$ is obtained and the iso-contour lines for different values of $g_t(\tilde{\rho}_t)$ are shown in Fig. 10a. The plotted lines are qualitatively similar to the experimental iso-contour lines shown in reference [8], for different $g_t(r_t)$ values when plotted between $d_{t,1}/d_p$ and $z_1/d_p$ (see Fig. 10b). Each iso-contour line at a given $g_t$ value represents a calibration curve, similar to the experimental calibration curve shown in Fig. 4. To quantitatively compare the theoretical and experimental calibration curves, the x-axis of the theoretical curve should be transformed from $\tilde{\sigma}$ = $\sigma/d_o$ to $\Delta z/d_o$. Here $\Delta z$ represents the particle distance from the object plane. Eq. (7) suggests a transformation relation between $\tilde{\sigma}$ = $\sigma/d_o$ and $\Delta z/d_o$, provided the experimental constant $\beta$ is known.

However, the estimation of $\beta$ by strictly theoretical means is not trivial, as shown by the flow chart in Fig. 11, which presents two procedures that could

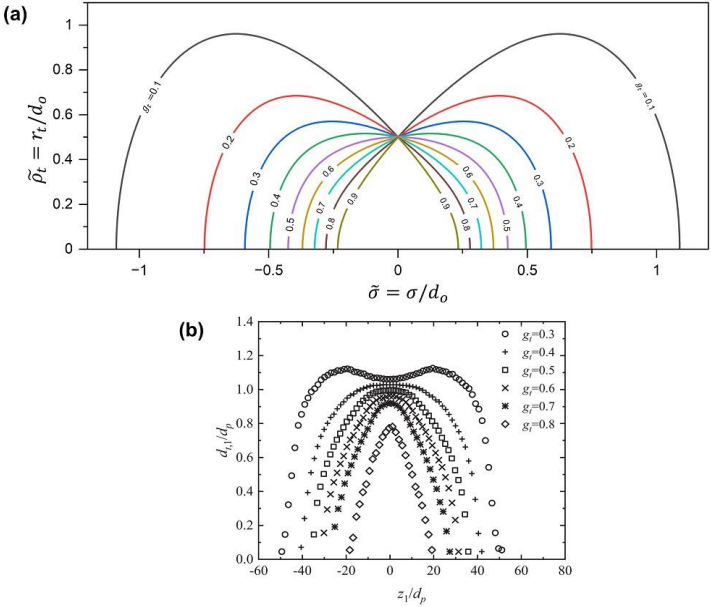

**Fig. 10** (a) Theoretical prediction for $\tilde{\rho}_t$ = $r_t/d_o$ variation with $\tilde{\sigma}$ = $\sigma/d_o$ for different intensity threshold values ($g_t$ = 0.1 to 0.9). (b) Similar experimental curve between $d_{t,1}/d_p$ vs $z_1/d_p$ from Zhou et al. 2020 [8].

be employed for obtaining the transformation relation between $\tilde{\sigma}$ = $\sigma/d_o$ and $\Delta z/d_o$. Method 1 involves solving a complex optimisation problem that does not require any experimental calibration for evaluating the transformation relation, i.e. a self-calibration. To obtain the calibration curve from Eq. (18) or through iso-contour $g_t$ plots (Fig. 10), the values of four unknowns i.e. $\beta$, $\sigma_1$, $\sigma_2$ and $d_p$ (note $d_o$ = M ×$d_p$) must be evaluated. To evaluate these unknowns, the value of $g_{t1}$ (intensity on camera 1) can be fixed arbitrarily, i.e. by choosing one of the iso-contour curves. Now, for an arbitrary guessed value of $d_p$ and a known value of $r_t$, which equals the detected size of an unknown particle on the image screen after thresholding at the chosen $g_{t1}$ level, $\tilde{\sigma}$ is evaluated from the chosen iso-contour curve from Fig. 10. This $\tilde{\sigma}$ value is used to evaluate the value of $\sigma_1$ for the corresponding value of $d_p$. In the next step, an intensity iso-contour line on camera 2 ($g_{t2}$) is chosen and the value of $r_{t2}$ for the same particle as done in step 1 is evaluated. Here also, a value of $\sigma_2$ can be obtained for each corresponding guessed value of $d_p$ by using the iso-contour intensity curve values $g_{t2}$. The following expression can be obtained using Eq. (7)

$$\sigma_1 - \sigma_2 = \beta(z_1 - z_2) \qquad (19)$$

Here, $z_1$ and $z_2$ are the object plane distances from the imaging lens, as shown in Fig. 4. Only one combination of $\sigma_1$ and $\sigma_2$ for a particular $d_p$ value will satisfy the Eq. (19) if $\beta$ remains constant throughout all measurements taken by the system. Thus, Method 1 represents a rather complex optimisation solution.

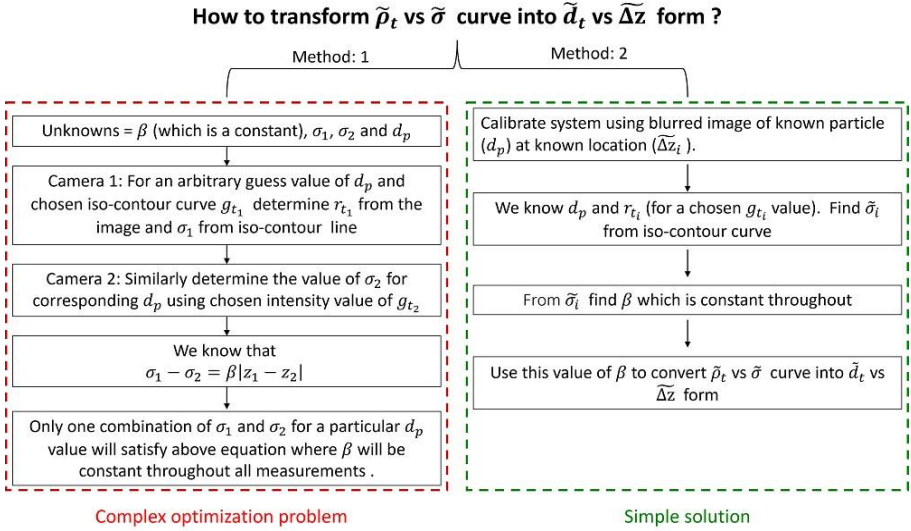

**Fig. 11** Flowchart for converting $\tilde{\rho}_t$ vs $\tilde{\sigma}$ curve into $\tilde{d}_t$ vs $\tilde{z}$ through two methods, optimisation based solution (Method 1) and simplified solution using two-point calibration (Method 2).

A simpler solution can be obtained by taking a single blurred image of a known target dot at a known position from the object plane. This algorithm is denoted as Method 2 on the processing flow chart shown in Fig. 11. The blurred image will provide an estimate for $r_{ti}$ for a chosen image intensity value $g_{ti}$. Here, subscript $i = 1,2$ corresponds to camera 1 and camera 2, respectively. As the size of the target dot ($d_t = M \times d_p$) is already known, for a known value of $\tilde{\rho}_{ti}$, the value of $\sigma_i$ can be obtained using Fig. 10a for the chosen $g_{ti}$ values. The experimental constant $\beta$ is obtained after substituting the value of $\sigma_i$ into Eq. (19). After replacing the $\beta$ value in the transformation Eq. (7), the calibration curve between $\tilde{\rho}_t$ and $\tilde{\sigma}$ is transformed into the $d_t/d_o$ vs. $\Delta z/d_o$ form. The transformed theoretical relation with comparison to experimental calibration curve is shown in Fig. 12. Good agreement is observed between theoretical and experimental curves, indicating that the Method 2 could potentially be utilised to provide a simplified approach for obtaining calibration function. However, due to some discrepancies which are observed for $\tilde{z} > 4$ values, only the experimental calibration curve has been used for the measurements presented in this study. The possible reason for such discrepancies and its rectification is yet unclear and an improved form of the theoretical calibration curve will be the subject of future work.

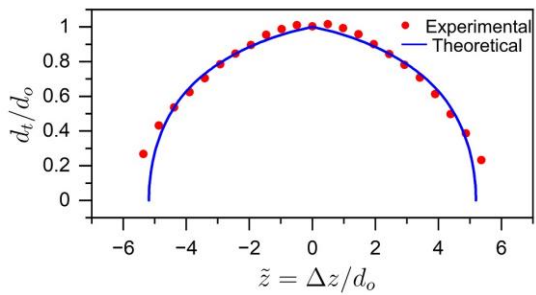

**Fig. 12** Theoretical vs experimental calibration curves using Method 2 shown in Fig. 11

## 4 Results and discussion

In this section measurement results will be shown for the atomization of a distilled water drop at three different Weber numbers of 900, 2000 and 4000. Measurements were performed 30mm downstream and 4mm below the initial droplet position, as indicated in Fig. 2a. The Weber number values correspond to the shear induced entrainment (SIE) mode of droplet breakup [6, 7, 26], where the initial drop undergoes fragmentation due to shearing of droplet fluid by its interaction with the shock-induced airflow. The external flow shears the liquid mass towards the droplet periphery by means of Kelvin-Helmholtz surface waves, which at a later stage accumulate on the droplet periphery to form a liquid sheet. The fragmentation of this liquid sheet accounts primarily for the generation of secondary droplets, whose size distribution is being measured in the present work. The mechanistic insights into the droplet breakup through SIE mode can be found in [7, 9] and is not the focus of present work. Figures 2a and b show the global overview of interaction dynamics for Weber number values of 879 and 4088, respectively. As can be seen from these figures, a dense mist of secondary droplets is produced during the breakup process and the intensity of atomization increases with increase in Weber number. Such interaction results in a dense spray of micron sized daughter droplets, the size distribution of which evolves over time. This then requires high spatiotemporal resolution of the measuring system, an ideal application of the DFD technique.

The first result is shown in Fig. 13, in which the measured number density distributions and volume weighted distributions are shown for the three Weber numbers. These distributions have been accumulated over the entire atomization duration, typically involving 15 experimental runs for each Weber number case. These distributions have been corrected for the differing detection volume as a function of drop size. The inset of these figures shows the values of $D_{v,10}$, $D_{v,50}$ and $D_{v,90}$ values, which indicate the droplet size below which 10%, 50% and 90 % of sample droplet volume lies, respectively. Note that the drop size distribution shifts to lower values with increasing Weber number, which is an expected and accepted dependency. At higher Weber numbers the number of smaller drops also increase proportionally. The volume weighted distributions indicate clearly that at low Weber number the larger drops carry a larger portion of the total bulk liquid mass. With increasing Weber number a larger portion of the liquid mass is found in small droplets.

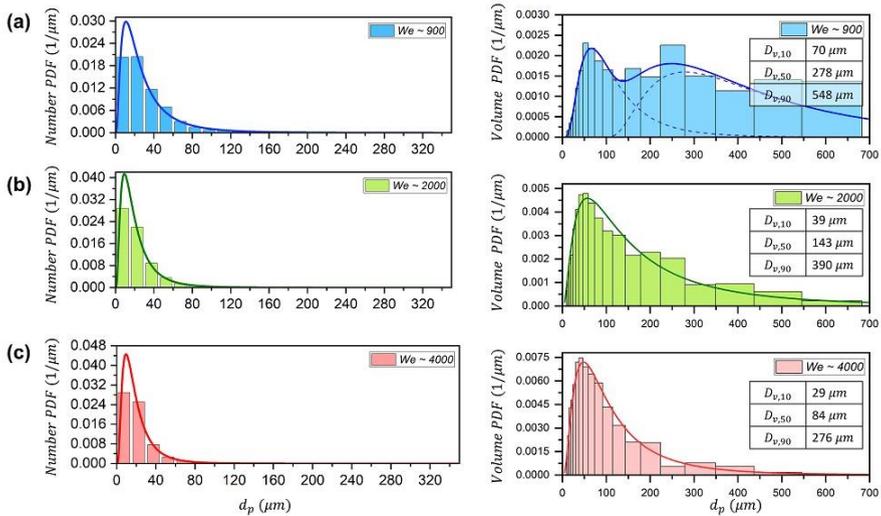

**Fig. 13** Number probability distribution (left column) and volume probability distribution (right column) of secondary droplets for (a) We ∼ 900, (b) We ∼ 2000, and (c) We ∼ 4000. The solid lines indicate the probability distribution function as per fitted log-normal function. For We ∼ 900 a multi-modal distribution is used.

Taking advantage of the high-speed cameras, the distributions shown above were then recomputed using three different time segments. For this the normalised time $\tau$ was introduced, whereby $\tau = 0$ corresponds to the first instant of daughter droplets appearing in the measurement region and $\tau = 1$ corresponds to the time when no further secondary drops are detected. The respective number distributions and volume distribution are shown in Figs. 14 and 15. As can be observed from these distributions, they exhibit a higher scatter, attributed to the fewer drops available for the computation of each distribution. The results show that secondary atomization is a time-varying process where the size distribution of the atomized droplets varies over time.

For the initial time period ($\tau = 0$–$0.33$), the distribution is less scattered, and droplets of comparatively smaller sizes are observed. This is expected, since the relative velocity between the external gas flow and the droplet is maximum during the initial interaction stage (decreases with time), resulting in a higher aerodynamic force acting on the droplet surface. Second, the smaller daughter droplets will follow the gas flow more faithfully (lower Stokes number) than the larger droplets, reaching more rapidly the measurement zone. It has also been shown in previous work [7] that the droplets in the initial stage are formed by the fragmentation of thin liquid sheets formed at the droplet periphery due to the accumulation of droplet fluid transported by the Kelvin-Helmholtz surface waves. This thin sheet fragmentation results in the generation of smaller daughter droplets. In contrast, at later instances a thicker liquid sheet is formed at the droplet periphery, due to the contribution of both Kelvin-Helmholtz wave based transportation and internal flow towards the periphery, arising from the droplet deformation. Thus, a thicker liquid sheet will fragment into comparatively larger daughter droplets, as shown in Figs. 14 and 15 for $\tau$ = 0.67 to 1. Lower aerodynamic forces further favour larger droplets during later time periods and higher Stokes number of the larger droplets, resulting in the increased time taken by the daughter droplets to reach the measurement region.

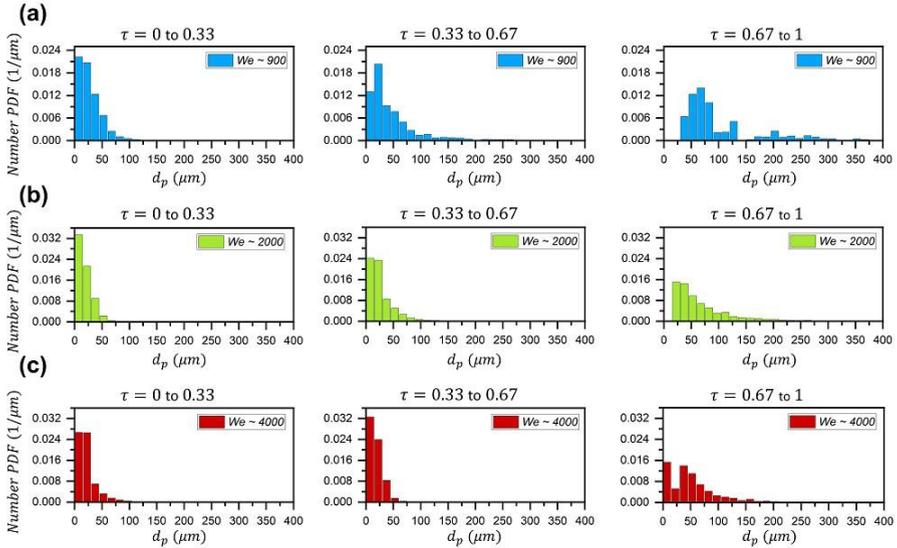

**Fig. 14** Number probability distribution variation with time for daughter droplets at (a) We ~ 900 (b) We ~ 2000 and (c) We ~ 4000.

From these distributions, either overall (Fig. 13) or time resolved (Figs. 14 and 15), it is then possible to compute the Sauter Mean Diameter (SMD), defined as

$$D_{32} = \frac{\sum_{i=1}^{i=N} n_i D_i^3}{\sum_{i=1}^{i=N} n_i D_i^2} \qquad (20)$$

where $N$ is the total number of drops in the respective distribution, $n_i$ is the number of drops in the $i^{th}$ bin and $D_i$ is the diameter in the middle of the $i^{th}$ bin. The SMD is a measure especially suited to characterise spray drop distributions when evaporation or solidification is involved, as in the present application of interest.

The SMD values computed for the entire atomisation process is shown in Fig. 16a in which a clear decrease in SMD is observed for increasing Weber number. Of particular interest is the dependency of SMD as a function of normalised time, shown in Fig. 16b. Here the SMD increases as the atomisation process proceeds, independent of which Weber number is considered.

These results indicate that the time and spatial resolution afforded by the depth from defocus imaging technique is particularly important in lending insight into the time-dependent atomization mechanisms encountered in this shock-drop interaction. In formulating a model for this phenomenon it is appropriate to cast this in terms of Weber and/or Mach number and these results make apparent the necessity to also include the evolution of the secondary atomization over time. This is not an entirely new insight, as similar observations have been made in [27]; however, the DFD technique, as implemented in this study, offers such information in more arbitrary spray situations.

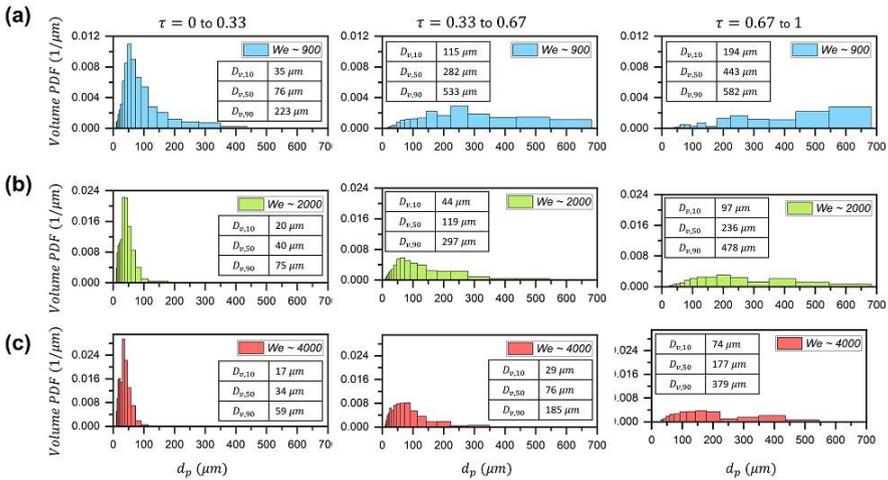

**Fig. 15** Volume probability distribution variation with time for daughter droplets at (a) $We \sim 900$ (b) $We \sim 2000$ and (c) $We \sim 4000$.

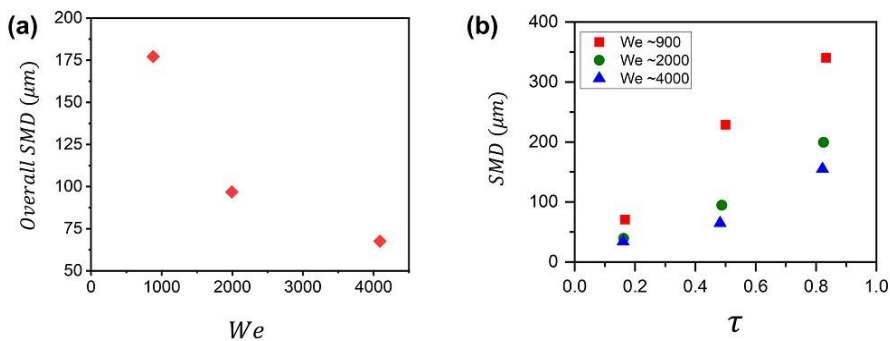

**Fig. 16** (a) Overall Sauter Mean Diameter ($D_{SMD}$) of daughter droplets for different Weber number (*We*) values. (b) SMD variation at different time instances for $We \sim$ 900, 2000 and 4000.

## 5 Conclusions

The camera depth from defocus (DFD) technique was implemented to perform drop size measurements in a shock-drop interaction. In particular, the shear Induced Entrainment (SIE) breakup mode was investigated, which results in a transient dense mist of micron size atomized droplets. The ability of the DFD technique to provide the high spatio-temporal resolution required for sizing such a dense spray phenomenon was effectively demonstrated. The measurements were corrected for a drop size dependent detection volume, resulting in a significant shift of size distributions toward the smaller drop sizes. This is one of the key improvements implemented in the present work and presently, there is no alternative measurement technique for drops in a spray that offers comparable accuracy in estimating the detection volume size; hence, measurements of the number density. Furthermore, a simplified and improved calibration procedure based on a theoretical model was introduced and validated with experimental data. These further developments of the DFD technique make it now a very attractive alternative to more established techniques for spray drop sizing, offering a simple optical arrangements, ease of use and very reasonable computational demands in the signal and data processing.

## Author contributions

Sharma - investigation, software, writing - original draft: Chandra - investigation, software; Jatin - investigation, software; Jatin and Chandra contributed equally; Basu - funding acquisition, project administration, supervision, conceptualization; Kumar - project administration, supervision; Tropea methodology, writing - original draft, writing - review & editing

# Acknowledgements

The financial support of this research by IGSTC (Indo–German Science and Technology Center) through project No. SP/IGSTC-18-0003 is thankfully acknowledged, as is the support of the Science and Engineering Research Board of India in sponsoring author CT through the VAJRA Faculty scheme. NKC acknowledges the support from Prime Minister's Research Fellowship (PMRF). A special thanks goes to Prof. Wu Zhou at the University of Shanghai Science and Technology for making a preliminary version of our image processing code available.

# Declarations

The authors report no conflict of interest.